\documentclass[lettersize,journal]{IEEEtran}
\usepackage{amsmath,amsfonts}
\usepackage{algorithmic}
\usepackage{algorithm}
\usepackage{array}
\usepackage[caption=false,font=normalsize,labelfont=sf,textfont=sf]{subfig}
\usepackage{textcomp}
\usepackage{stfloats}
\usepackage{url}
\usepackage{verbatim}
\usepackage{graphicx}
\usepackage{cite}
\usepackage{tabularx}
\usepackage[dvipsnames]{xcolor}
\usepackage{hyperref} 
\hypersetup{colorlinks=true, linkcolor=blue, citecolor=blue, urlcolor=blue}

\begin{document}

\title{A Survey on Advancements in THz Technology for 6G: Systems, Circuits, Antennas, and Experiments}


\author{Sidharth~Thomas,~\IEEEmembership{Graduate Student Member,~IEEE}, 
        Jaskirat~Singh~Virdi,~\IEEEmembership{Graduate Student Member,~IEEE}, 
         Aydin~Babakhani,~\IEEEmembership{Senior Member,~IEEE}, Ian~P.~Roberts,~\IEEEmembership{Member,~IEEE}
\thanks{Sidharth Thomas, Jaskirat Singh Virdi, Aydin Babakhani, and Ian P. Roberts are with  the Department of Electrical Engineering, University
of California, Los Angeles, CA 90095 USA (e-mail: sidhthomas@ucla.edu).}
}



\maketitle

\begin{abstract}
Terahertz (THz) carrier frequencies (100~GHz to 10~THz) have been touted as a source for unprecedented wireless connectivity and high-precision sensing, courtesy of their wide bandwidth availability and small wavelengths, but noteworthy implementation challenges remain to make this a reality.
In this paper, we survey recent advancements in THz technology and its role in future 6G wireless networks, with a particular emphasis on the 200--400~GHz frequency range and the IEEE 802.15.3d standard.
We provide a comprehensive overview of THz systems, circuits, device technology, and antennas, while also highlighting recent experimental demonstrations of THz technology. 
Throughout the paper, we review the state-of-the-art and call attention to open problems, future prospects, and areas of further improvement to fully realize the potential of THz communication in next-generation wireless connectivity.

\end{abstract}


\section{Introduction}
    \IEEEPARstart{W}{ireless} communication has become an indispensable component of contemporary life, with each generation of technological advancement opening the door to a host of novel applications that transform our lifestyles. 
With commercial 5G systems under deployment, the focus of many wireless researchers has shifted towards developing the sixth generation (6G) of wireless communication \cite{thz_rappaport, review_what_should_6G}. 
The forthcoming wave of technological advancement is poised to revolutionize wireless connectivity by offering unprecedented throughput and latency and expanded capabilities such as sensing, tailored to meet the stringent demands of applications of the next decade. 

\begin{figure}
        \centering
        \includegraphics[page = 7, trim=0.2cm 11.4cm 11cm 11cm, clip=true, width=0.45   \textwidth]{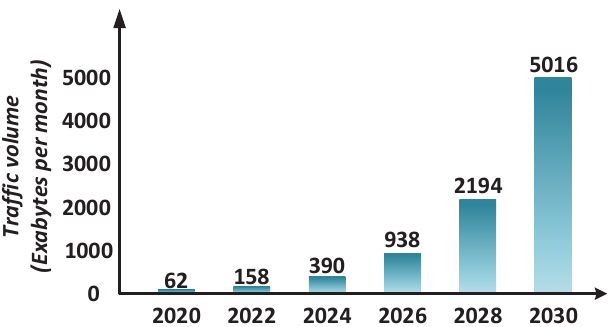}         
        \caption{The ITU's predicted growth in mobile data traffic through 2030 \cite{imt}. }
        \label{fig:data_curve}
 \end{figure}

The necessity for 6G technology primarily stems from the global increase in data generation and consumption \cite{cisco}. 
The amount of data traveling across the internet has exponentially grown over the years, and this trend is expected to continue.
The International Telecommunication Union (ITU) predicts that by 2030, global mobile traffic volume will be 670 times higher than in 2010, with aggregate mobile data traffic expected to exceed 5 ZB per month \cite{imt, 6g_survey_bs}. 
As illustrated in Fig.~\ref{fig:data_curve}, the traffic volume {per mobile device} in 2030 is expected to increase by 50 times from 2010. 
This surge is primarily attributed to the widespread proliferation of IoT (Internet of Things) devices, ranging from simple household gadgets to advanced industrial sensors, resulting in a massive and ever-growing volume of data. 
Automation technologies such as digital twins and the adoption of cloud-based machine learning and artificial intelligence across various sectors have further accelerated this trend \cite{review_towards}. 
As society's dependence on smart devices and connected systems increases, the demand for a more robust and capable network infrastructure becomes imperative \cite{jornet_nature_survey}. 
In response, 6G aims to deliver speeds, reliability, and latency improvements that far surpass those of its predecessors, establishing a critical foundation for the next digital connectivity and innovation era. 
The impressive wireless networks of today are the product of decades of extensive optimization and technology advancements. 
To deliver future 6G networks which substantially exceed the capabilities of today's 5G networks, many have proposed turning to vast, untapped swaths of spectrum at terahertz (THz) frequencies.


The THz band, which ranges from 0.1 to 10 THz \cite{athz_ref_1, athz_ref_2, athz_ref_3}, occupies a distinctive spot in the electromagnetic (EM) spectrum. 
Positioned between radio frequencies (RF) and optical frequencies, this band displays characteristics of both and offers exciting possibilities for new applications. 
The THz band can support high-capacity wireless links due to its vast bandwidth availability. 
Additionally, its short wavelength allows for operating large-scale antenna arrays in a small form factor. 
This facilitates network densification and makes THz ideal for sensing applications, with THz radars and imagers providing superior range and lateral resolution \cite{ehsan_imaging}. 
Moreover, THz imaging technologies may offer safer, more accessible medical diagnostics than traditional X-ray imaging \cite{tonouchi2007cutting}.
These unique features also enable future 6G technologies to combine communication and sensing into a single system, revolutionizing everyday interactions, transforming healthcare, and forging new markets.

A few THz bands in particular have garnered substantial interest for their applications in wireless communication. 
These bands include the D-band (110--170~GHz) \cite{dband_rebeiz, dband_nokia, dband_intel}, the 300~GHz band (253--322~GHz) \cite{fuji_1, fuji_2, fuji_3, okada_300, okada_2d, pfieffer}, and the 400~GHz band \cite{400g_thomas, reynart_outphasing}. 
Operating at these THz frequencies presents unique challenges spanning various technical considerations such as device technology, circuit design, antenna, packaging, channel modeling, signal processing, and system design \cite{isscc_6Gforum, 6g_survey_bs}.
These complexities associated with realizing THz transceivers and deploying THz wireless systems has drawn considerable attention and investment from both industrial and academic research laboratories.
Progress thus far has resulted in emerging THz wireless standards, with perhaps the most notable being IEEE 802.15.3d \cite{ieee_std}. 
This standard ultimately seeks to enable wireless communication over channels as wide as 69~GHz within the 253--322~GHz frequency range \cite{petrov2020ieee}. 
Its \textcolor{black}{aim} is to showcase the feasibility of THz frequencies for communication while also serving as a coordinated effort toward effective and reliable connectivity solutions at THz.

In this paper, we survey several key techniques and technologies for enabling wireless communication at carrier frequencies beyond 200~GHz and focus in particular on current specifications of the IEEE 802.15.3d standard. 
Our aim is to complement existing THz systems-level surveys (e.g., \cite{thz_rappaport, 6g_survey_bs, jornet_nature_survey, athz_ref_1, athz_ref_2, athz_ref_3, thz_survey_sp,thz_survey_ch, thz_survey_new}) by bridging topics at the intersection of multiple technical communities.
In this pursuit, we conduct a comprehensive analysis of the latest developments, potential opportunities, challenges, and the current state-of-the-art in the 200--400 GHz range. 
Unlike existing surveys on THz for 6G, this paper provides thorough discussion on semiconductor device technologies, circuit topologies, antennas, and packaging---and how design decisions surrounding such may impact system performance and capabilities.
We describe the unique challenges in these areas specific to THz design, acknowledge the limitations of current technology, and highlight promising approaches for practical solutions.
Furthermore, we overview notable demonstrations of THz-based 6G from across the globe and examine their architectural and implementation considerations, a crucial step to hone in on capable, cost-effective THz systems for future commercial 6G systems.



The rest of this paper is organized as follows. 
Section II \textcolor{black}{reviews} the envisioned role of THz in 6G. 
Section III describes the IEEE 802.15.3d standard. 
Section IV examines the behavior of wireless channels at THz frequencies. 
Section V \textcolor{black}{surveys the state-of-the-art} in semiconductor device technology. 
\textcolor{black}{Circuit techniques and architectures suited for THz design are discussed in Section VI.}
Antenna design and packaging techniques are addressed in Section VII. 
Section VIII showcases notable THz demonstrations worldwide. 
Finally, the paper is concluded in Section IX, summarizing the key findings and insights of this survey. 
Table \textcolor{black}{I summarizes key acronyms used throughout this paper}.

\begin{table}[t]
\centering 
\caption{List of Acronyms} 
\renewcommand{\arraystretch}{1.5} 
\begin{tabular}{|l|l|} 
\hline
5G & Fifth generation of wireless communication \\  \hline
6G & Sixth generation of wireless communication \\  \hline
BER & Bit error rate \\  \hline
BiCMOS & Bipolar CMOS \\  \hline
BPSK & Binary phase shift keying \\  \hline
CMOS & Complementary metal-oxide-semiconductor \\  \hline
DS & Directive scattering \\  \hline
EM & Electromagnetic \\  \hline
FDSOI & Fully depleted silicon on insulator \\  \hline
FSPL & Free-space path loss \\  \hline
$f_{\mathrm{max}}$ & Maximum oscillation frequency \\  \hline
$f_{\mathrm{T}}$ & Transit frequency \\  \hline
GaAs & Gallium arsenide \\  \hline
HBT & Heterojunction bipolar transistor \\  \hline
HEMT & High-electron-mobility transistor \\  \hline
IC & Integrated circuit \\  \hline
InP & Indium phosphide \\  \hline
IoT & Internet of things \\  \hline
ITU & International telecommunication union \\  \hline
JCAS & Joint communication and sensing \\  \hline
LNA & Low-noise amplifier \\  \hline
LO & Local oscillator \\  \hline
LDPC & Low-density parity-check code \\  \hline
$M$-QAM & $M$-ary Quadrature amplitude modulation \\  \hline
MIMO & Multiple input multiple output \\  \hline
OAM & Orbital angular momentum \\  \hline
OOK & On-off keying \\  \hline
OTA & Over-the-air \\  \hline
PA & Power amplifier \\  \hline
PCB & Printed circuit board \\  \hline
QPSK & Quadrature phase shift keying \\  \hline
RIS & Reconfigurable intelligent surface \\  \hline
SiGe & Silicon-germanium \\  \hline
SNR & Signal-to-noise ratio \\  \hline
THz & Terahertz \\  \hline
THz-OOK PHY & Terahertz on-off keying physical layer \\  \hline
THz-SC PHY & Terahertz single carrier physical layer \\  \hline
WLAN & Wireless local area network \\  \hline

\end{tabular}
\label{table:acroym_table} 
\end{table}

 \section{Applications of 6G: Why THz?}
    
\begin{figure}
        \centering
        \includegraphics[page = 8, trim=0.2cm 8cm 5.7cm 8cm, clip=true, width=0.45   \textwidth]{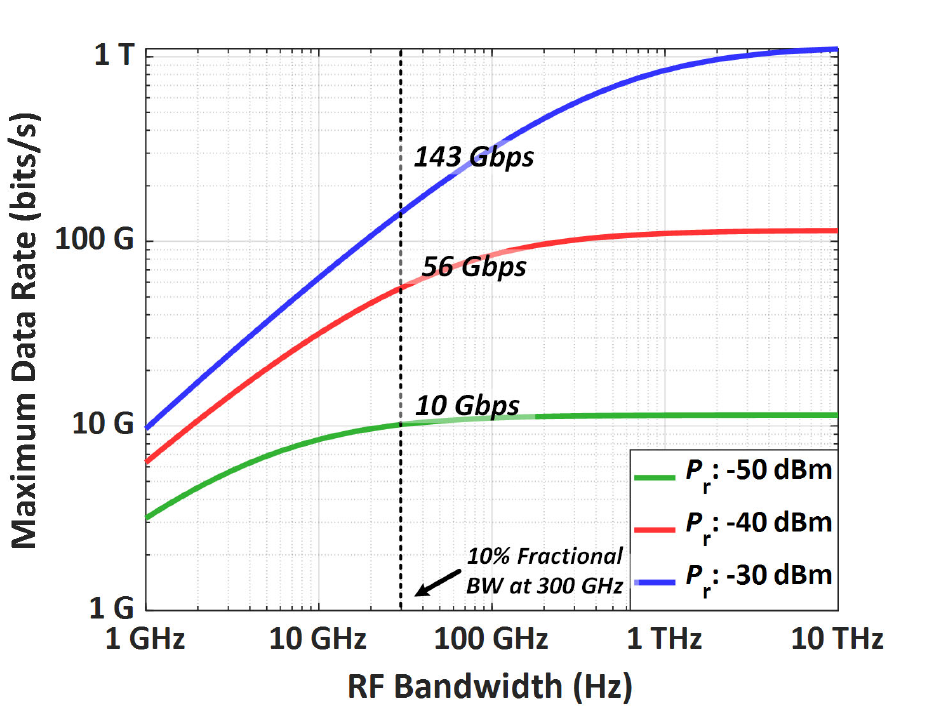}         
        \caption{The maximum data rate that can be achieved using the Shannon channel capacity theorem plotted against the RF bandwidth for different received powers, $P_\mathrm{r}$, assuming a receiver noise figure of 25 dB  \cite{isscc_6Gforum}. 
        The dashed line represents a typical scenario with 30 GHz RF bandwidth, which can be realized by having 10\% fractional bandwidth with a 300~GHz carrier. }
        \label{fig:capacity}
 \end{figure}

 The maximum data rate for a wireless channel with bandwidth \textcolor{black}{$B$} and a signal-to-noise ratio (SNR) of $\mathsf{SNR}$ can be calculated using the well-known Shannon channel capacity 
\begin{equation}\label{shannon}
R = B\cdot\log_2(1+\mathsf{SNR}).
\end{equation}
While this expression does not strictly depend on carrier frequency, bandwidth availability tends to increase at higher carrier frequencies \cite{goldsmith2005wireless,rappaport_5G}.
Increasing bandwidth $B$ then facilitates higher data rates, assuming appreciable SNRs can be attained.
For instance, at 300 GHz, the maximum data rate for link with a \textcolor{black}{$25$} dB noise figure (representative number \cite{isscc_6Gforum}) is plotted in Fig.~\ref{fig:capacity} for various received power levels. 
With a \textcolor{black}{$10\%$} fractional bandwidth (i.e., $B=30$ GHz) and \textcolor{black}{$-40$}~dBm received power, a link can support \textcolor{black}{$56$} Gbps. 
Such massive bandwidths also introduce the potential to multiplex an unprecedented number of devices across frequency resources, with each device enjoying substantial data rates.
To further illustrate the impact of enhanced throughput and latency offered by THz, this section highlights targeted applications conceptualized for THz-based 6G.

\begin{itemize}

   \item \textbf{WLAN and IoT:} THz-based 6G links have the potential to establish high-speed wireless local area network (WLAN) communication. 
   The main premise is that a THz access point serves as a sort of enhanced hotspot that can facilitate applications such as near-instant video transfer, large-scale IoT, and real-time streaming of immersive experiences in virtual reality \cite{review_white_paper, athz_ref_1}. As communication systems evolve and merge with IoT, 6G networks will connect humans, sensors, and computing devices across an extensive network \cite{6g_iot}.

  \item \textbf{Wireless backhaul:} With the evolution of wireless networks, mm-Wave small cells have become increasingly necessary to meet capacity demands. 
  A reliable and high-speed backhaul connection to each mm-Wave base station is essential to connect it to the core network, but traditional fiber backhauling does not scale well due to the sheer density of small cells \cite{cudak21integrated,polese2020integrated}.
  Wireless backhauling via THz could interconnect base stations to the core network with extremely high capacity links, circumventing the time, cost, and overhead associated with trenching optical fiber \cite{petrov2020ieee, jornet_km}.
  In turn, this could facilitate a paradigm where wireless networks provide computing and cloud-based services directly to devices at the network's edge, proliferating emerging technologies such as artificial intelligence (AI) and digital twins \cite{twins}.  

  \item \textbf{AR/VR and holographic projections:} Augmented reality (AR), virtual reality (VR), and 3D holographic projections are all application spaces where the impact of 6G technology is expected to be substantial \cite{thz_ref_5, thz_ref_9}. 
  In essence, providing a high-capacity, low-latency THz link can facilitate computational offloading from AR/VR headsets, allowing them to reduce their form factor and battery consumption, while remaining immersive and interactive. 
  Advancements in this area can even revolutionize the healthcare industry and potentially enable applications such as remote surgeries \cite{thz_ref_10}.

  \item \textbf{Kiosk downloading:} In this setup, a fixed wireless transmitter, unconstrained by size and power limitations, may connect to a fiber network, sending high-speed multi-Gbps wireless data to a low-power mobile receiver within a short range. This may be utilized to transfer massive files and high-definition videos in a ``touch-and-go'' fashion \cite{kiosk}. This application has been gaining attention with prototypes being demonstrated \cite{kiosk}. 
  
  \item \textbf{Data centers:} THz links have the potential to supplement or replace current \textcolor{black}{Ethernet and fiber-based} links in data centers \cite{petrov2020ieee}.
  Current approaches require extensive cables, which can be expensive and difficult to set up.
  An alternative solution is utilizing point-to-point high-speed wireless links connecting racks. 
  This is more flexible and could reduce the cost of cabling. 
  \textcolor{black}{Moreover, this approach reduces latency as the EM waves can travel through air over 50\% faster than through optical fiber \cite{pozar}}.

  \item \textbf{Joint communication and sensing:} 
  A so-called joint communication and sensing (JCAS) paradigm in which wireless communication systems also perform sensing tasks has gained traction as a defining feature of future 6G networks \cite{jcas}.
  The main premise behind JCAS is to leverage existing communications infrastructure, such as cellular network base stations, to perform wide area sensing, as opposed to deploying dedicated sensing hardware (e.g., radar).
  In doing so, emerging applications which rely on sensing data, such as real-time digital twins and autonomous driving, could be fully realized. 
  Simultaneously, this sensing data could actually be leveraged to improve communication system performance by, for example, adjusting system operation dynamically upon detecting blockage/obstructions, such as vehicles \cite{Ericsson2021JCAS}.
  Beyond this, applications of JCAS are ripe in industrial automation, robotics, and AR/VR, especially in environments/conditions where camera may fail due to darkness and/or weather.
  With THz, the potential for JCAS is particularly exciting, as wide bandwidths yield fine range resolution, dense antenna arrays provide granular angular resolution, and high carrier frequencies facilitate Doppler-based velocity estimation.

 
  \item \textbf{Secure communication:}  THz links have emerged as a promising solution for enhancing communication security. 
  With their highly directional nature, eavesdropping becomes significantly more difficult, offering better data protection and privacy \cite{jornet_nature_survey}. 
  The high levels of atmospheric attenuation also enable new ideas such as `whisper radio,' where communication is inherently secure \cite{scatter_rappaport}.
  Additionally, the small wavelength of THz signals, which measures in the range of hundreds of micrometers, allows for innovative modulation techniques, which can enhance physical layer security. One example is in space-time modulated phased arrays \cite{sengupta_Secure}. This is illustrated in Fig.~\ref{fig:secure}\textcolor{blue}{(a)}. Here, parallel data streams are modulated spatially at different transmitters and temporally at different instances. This data can be deciphered by the intended broadside receiver, Bob. \textcolor{black}{However, an eavesdropper in the off-axis (such as Eve) cannot decipher the information, as the signals from each antenna travel different distances, resulting in signal corruption.} Another example is using orbital angular momentum (OAM), for wireless communication \cite{han_oam, oam_wei}. Here, the information is modulated on a carrier signal, which has an angular momentum mode, as illustrated in Fig.~\ref{fig:secure}\textcolor{blue}{(b)}. A receiver antenna can only decipher the information if it has the same angular momentum mode as the transmitter.

 \begin{figure}
        \centering
        \includegraphics[page = 12, trim=0.2cm 8cm 7.6cm 8.4cm, clip=true, width=0.45   \textwidth]{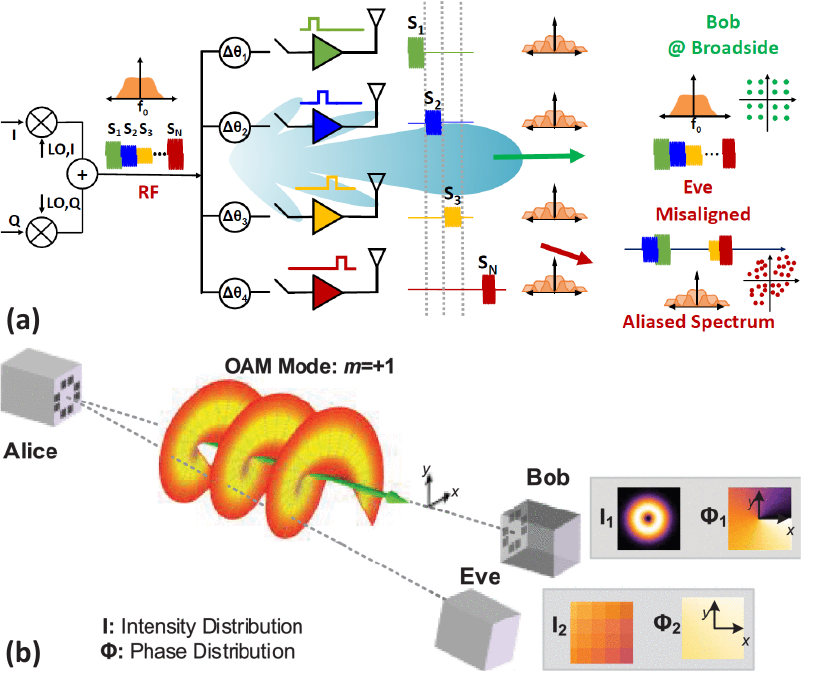}         
        \caption{(a) Application of a space-time modulated phased arrays in transmitting data selectively towards the intended receiver, Bob, while corrupting the data constellation and spectrum towards the eavesdropper, Eve \cite{sengupta_Secure}. (b) A THz OAM transceiver can establish a physically secure wireless link that is resistant to off-beam-axis eavesdropping. Here, Bob, the intended receiver, operates in the same mode as Alice and can receive the information. The eavesdropper, Eve, is off-axis and cannot receive information \cite{han_oam}.  }
        \label{fig:secure}
 \end{figure}


  \item \textbf{Satellite communication:} THz links may in fact also offer a viable option for ground-satellite connections, since atmospheric attenuation decreases exponentially with altitude \cite{satellite}. 
  As an example, at 600 GHz, the free-space path loss of a geostationary link originating from a moderately dry, high-altitude location is equivalent to that of a 2-km link situated at sea level.
  THz links can also be used for inter-satellite communication, with the potential to outperform optical links commonly used today \cite{satellite}. 
  The vacuum of space eliminates atmospheric attenuation issues, and the larger THz wavelengths (compared to optics) result in much broader beam widths, which enhances the accuracy between the transmitter and the receiver. 
  Additionally, THz-band communication systems often require significantly less power and weight than their laser-based counterparts \cite{space, isscc_6Gforum}. 

    
\end{itemize} 

\section{\textcolor{black}{The} IEEE 802.15.3d Standard}
     The IEEE 802.15 Terahertz Interest Group was established in 2008 to explore wireless communication within 0.3 to 3 THz. In 2017, the IEEE 802.15.3d standard (the 300~GHz band) was approved, serving as the initial step towards THz band communication \cite{ieee_std}. This standard outlines a PHY layer of 100 Gbps, with the option to revert to lower rates. 
 It supports wireless communications over up to 69 GHz wide channels within 253–322 GHz. Its primary goal is to demonstrate the practicality of using THz frequencies for communication while providing high-connectivity solutions. 
 Note that the envisioned applications are restricted primarily to point-to-point links between static or quasi-static devices, to remain within the realm of what is possible with current semiconductor technology. 
 However, this may change in the future as device technology matures.
 A detailed description of this standard can be found in \cite{petrov2020ieee}.

 \begin{figure}
        \centering
        \includegraphics[page = 9, trim=0.2cm 8cm 8cm 9.5cm, clip=true, width=0.45   \textwidth]{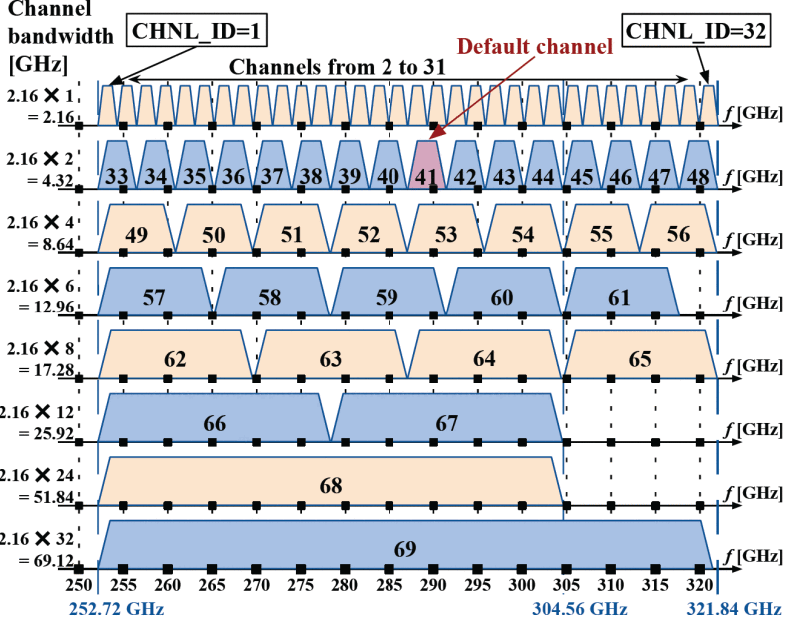}         
        \caption{ \textcolor{black}{Channel plan for the IEEE 802.15.3d standard \cite{petrov2020ieee}. There are 69 overlapping channels between 252.72 GHz and 321.84 GHz, which support 8 channel bandwidths from 2.16 GHz up to  69 GHz.}     
        }
        \label{fig:channels}
 \end{figure}
 
As illustrated in Fig.~\ref{fig:channels}, the IEEE 802.15.3d  standard covers frequencies between 252.72 GHz and 321.84 GHz, with 69 overlapping channels. 
These channels offer eight supported bandwidth options, ranging from 2.16 GHz to 69.12 GHz, each bandwidth being an integer multiple of 2.16 GHz. 
By default, channel number 41, with a bandwidth of 4.32 GHz, is used. 
Summarized next, the PHY layer in the IEEE 802.15.3d standard has two modes: THz single-carrier mode (THz-SC PHY) and THz on-off keying mode (THz-OOK PHY).
 
 \subsection{THz-SC PHY: \textcolor{black}{Single-Carrier Mode}}
THz-SC PHY aims to achieve high data rates and caters to bandwidth-oriented use cases such as wireless fronthaul/backhaul and additional data center links. The mode offers a range of six different modulations, including four variations of phase-shift keying: binary (BPSK), quadrature (QPSK), 8-phase (8-PSK), and 8-phase asymmetric (8-APSK). In addition, quadrature amplitude modulation \textcolor{black}{(QAM)} is available as 16-QAM and 64-QAM. While BPSK and QPSK modulations are mandated for the THz-SC mode, the other modulations are optional. 
This mode employs one of two low-density parity check (LDPC) codes for forward error correction: a high-rate 14/15 LDPC (1440, 1344) or a low-rate 11/15 LDPC (1440, 1056).

 \subsection{THz-OOK PHY: \textcolor{black}{On-Off Keying Mode}}
THz-OOK PHY mode is designed to cater to \textcolor{black}{low-complexity THz devices by using low-cost, relatively simple amplitude modulation schemes}. 
Despite this limitation, it can still attain impressive \textcolor{black}{data} rates of up to tens of gigabits per second when utilizing the widest channels available. 
In terms of coding, three different error correction schemes are supported, including the mandatory (240, 224)-Reed Solomon code for simple hard decoding. 
Additionally, the two LDPC-based schemes described in the the previous subsection can also be used with THz-OOK operation for soft decoding.

 \section{Wireless Channel Characteristics 
 }
    Wireless channels at THz bands exhibit behaviors markedly distinct from other key frequencies, \textcolor{black}{such as the sub-6 GHz band and mm-Wave bands}. 
This difference arises from the unique interaction between EM waves and materials at THz frequencies, whose dynamics are heavily dependent on the size of the wavelength relative to the physical dimensions of the objects it interacts with. 
THz channels show high levels of signal attenuation and display scattering properties notably different from their lower-frequency counterparts. 
Moreover, there is strong absorption loss, \textcolor{black}{susceptibility to blockage, and shadowing}  \cite{thz_rappaport, review_what_should_6G}. 
This results in a sparse channel with few paths from the transmitter to the receiver \textcolor{black}{and can lead to reliability concerns in certain environments}. 

 \begin{figure}
        \centering
        \includegraphics[page = 1, trim=0cm 8.6cm 0.6cm 9cm, clip=true, width=0.45   \textwidth]{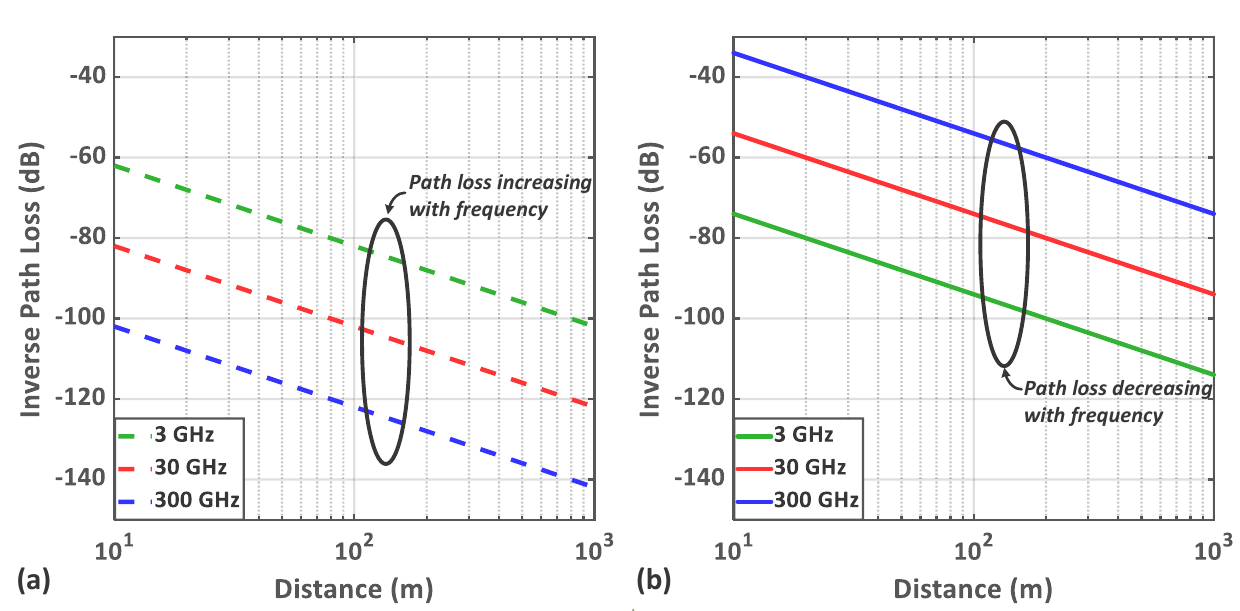}         
        \caption{Inverse path loss, $(1/ L)$, versus distance for different frequencies. Note that the antenna gain is embedded with the loss. (a) The dashed line represents the loss with an isotropic antenna. The path loss increases with frequency for a given distance. (b) The solid line represents the loss with a directive antenna (having a fixed aperture for all three frequencies). The path loss decreases with frequency for a given distance. 
        }
        \label{fig:fspl}
 \end{figure}

\subsection{Path Loss}

Path loss in wireless communication primarily stems from free-space path loss (FSPL) and atmospheric attenuation, each characterized by distinct mechanisms as outlined below.

\subsubsection{Free-Space Path Loss}
FSPL occurs from the radial spreading of EM waves as they travel through free space. This effect is quantified by (\ref{friss}) 
\begin{equation}\label{friss}
L = \left(\frac{4 \pi d}{\lambda}\right)^2 ,
\end{equation}
where $L$ represents the path loss, $d$ signifies the distance the wave has propagated, and $\lambda$ is the wavelength of the EM wave \cite{pozar}. 
Note that the antenna gain can be embedded along with FSPL. 
The FSPL with isotropic antennas at 3 GHz, 30 GHz, and 300~GHz are plotted with dashed lines in Fig.~\ref{fig:fspl}\textcolor{blue}{(a)}. 
It can be observed that the path loss increases with frequency.

\subsubsection{Atmospheric Attenuation}
\begin{figure}
        \centering
        \includegraphics[page = 2, trim=0cm 8cm 5cm 8cm, clip=true, width=0.4   \textwidth]{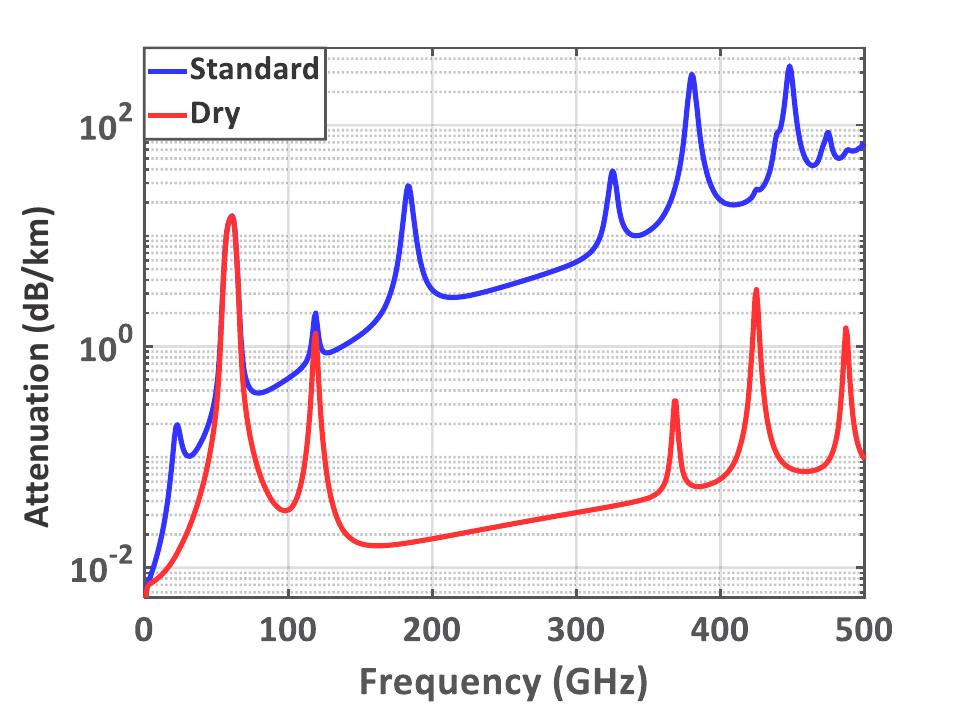}         
        \caption{ Atmospheric attenuation versus frequency in standard (blue) and dry (red) conditions \cite{attenuation}. Strong absorption bands can be observed at 183 GHz, 325 GHz, 380 GHz, and 450 GHz. }
        \label{fig:atmospheric_attenuation}
 \end{figure}

THz frequencies experience higher atmospheric attenuation than RF frequencies due to various effects. THz wavelengths are comparable in size to dust, rain, snow, and other atmospheric particles, which leads to higher attenuation \cite{thz_rappaport}. Moreover, several absorption resonances exist at THz frequencies. 
This happens when the frequency approaches the energy required to excite vibrational modes in molecules. 
The THz absorption spectrum is plotted in Fig.~\ref{fig:atmospheric_attenuation} for standard (blue) and dry (red) atmospheric moisture conditions. 
Sharp absorption peaks are observed at 183~GHz, 325~GHz, 380~GHz, and 450~GHz, which arise due to rotational and vibrational excitation states in gas molecules \cite{attenuation}. 
While these bands have some applications in short-range secure communication schemes such as `whisper radio' \cite{scatter_rappaport}, and in applications such as hydration sensing and spectroscopy \cite{hitran, thz_spectro}, they should be avoided for long-distance communication. 
Note that some frequency bands, such as the D-band (110-170~GHz), 300~GHz, and 400~GHz bands, do not have significant absorption peaks. Hence, these frequencies are optimal candidates for THz-based 6G communication. 
At 300~GHz, atmospheric attenuation is as low as 10 dB per km, making it tolerable for communication links \cite{attenuation, thz_rappaport}.

  \begin{figure}
        \centering
        \includegraphics[page = 3, trim=0cm 8cm 0cm 11cm, clip=true, width=0.48   \textwidth]{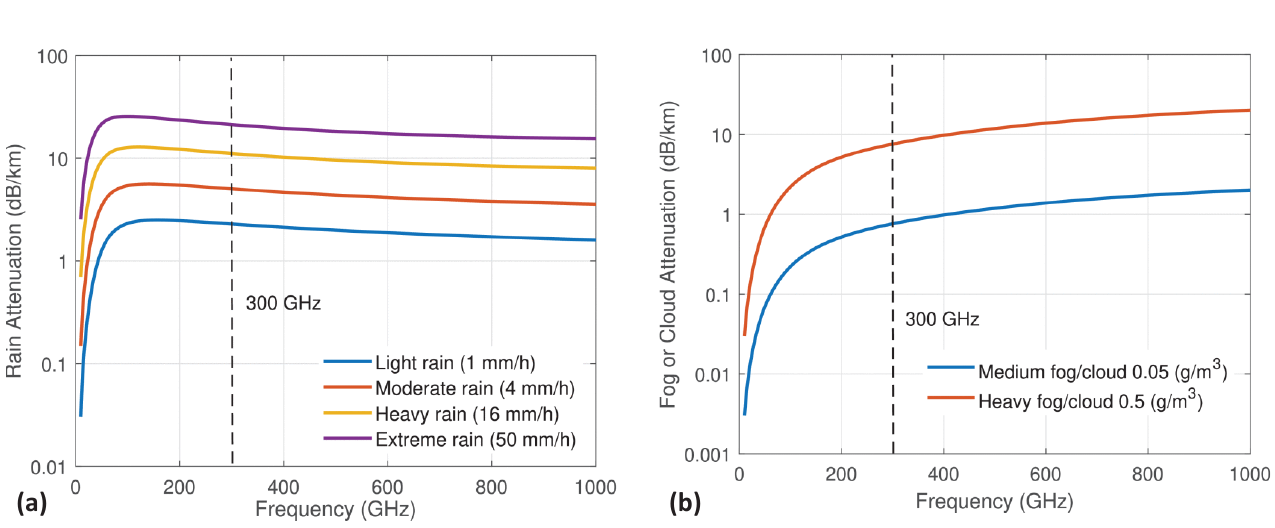}         
        \caption{(a) Measured rain attenuation for horizontal polarization. Attenuation due to rain saturates beyond 100 GHz. (b) Measured fog/cloud attenuation, which slowly increases with frequency. \cite{channel_survey}  }
        \label{fig:rain_fog}
 \end{figure}

THz links are also prone to adverse weather conditions such as rain, snow, hail, fog, and clouds. This has been extensively studied, and it has been shown that rain attenuation flattens out after 100 GHz \cite{rain_attenuation, rain, snow} (Fig.~\ref{fig:rain_fog}\textcolor{blue}{(a)}). 
At 100 GHz, the attenuation caused by fog and cloud is approximately 5 dB per kilometer. 
In the most severe conditions of heavy fog or cloud, it can increase by roughly 15 dB when the frequency rises from 0.1 to 1 THz. 
This is plotted in Fig.~\ref{fig:rain_fog}\textcolor{blue}{(b)} \cite{channel_survey}. This additional attenuation may be compensated by using high-gain antennas.

\subsection{Scattering}
The scattering behavior of an EM wave with a material depends on the roughness of the scatterer with respect to the wavelength \cite{scatter_rappaport}. The Rayleigh criterion can be used to determine the smoothness or roughness of a surface \cite{rappaport2024wireless}. It defines a critical surface height of a surface, $h$, which depends on the wavelength and the angle of incidence, $\theta_i$. The critical height is given in (\ref{scatter}). 

\begin{equation}\label{scatter}
h = \frac{\lambda}{8\cos \theta_i}
\end{equation}

\begin{figure}
        \centering
        \includegraphics[page = 4, trim=0cm 8cm 13cm 7.8cm, clip=true, width=0.45   \textwidth]{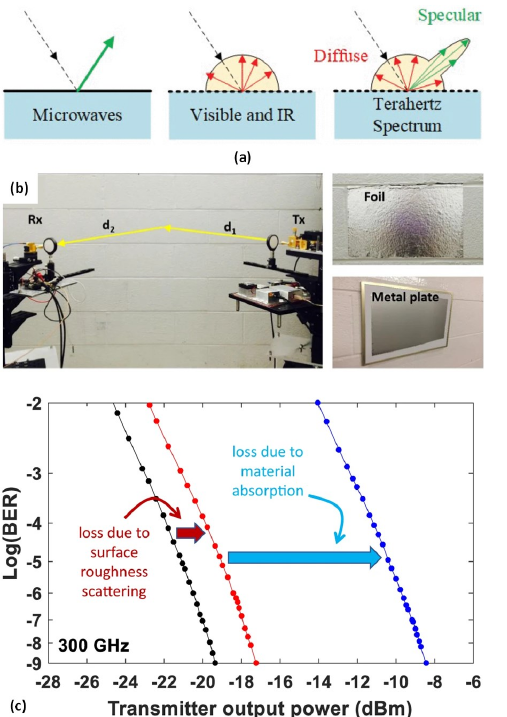}         
        \caption{(a) Surface scattering exhibiting specular behavior for microwaves, diffuse behavior for visible and infrared frequencies, and a combination of both for THz frequencies \cite{thz_rappaport} (b) Setup to measure bit error rate (BER) for a 2-m link involving reflection from a cinderblock wall, reflection from the wall coated with metal foil, and reflection from a metal plate, and (c) measured BER for the three cases \cite{scatter_meas}. Here, the black curve denotes reflection from a metal plate, the red curve denotes reflection from the wall coated with metal foil, and the blue curve denotes reflection from the cinderblock wall. }
        \label{fig:scattering}
 \end{figure}

The surface is smooth for a given frequency with wavelength $\lambda$ if the minimum to the maximum surface protuberance, $h_0$, is smaller than $h$. At RF frequencies, most surfaces are smooth and follow reflection-like specular behavior. Here, the reflection process has a strong dominant specular path, where the incident angle equals the reflected angle. There is also little absorption. This results in a rich multi-path channel that can be studied with several models. At optical frequencies, the scattering is diffuse, with multiple signal paths across different directions (Fig.~\ref{fig:scattering}\textcolor{blue}{(a)}). 

The channel behavior is much different at THz. Here, the surface roughness becomes comparable to the wavelength, and hence, scattering shows both diffuse and specular behavior (Fig.~\ref{fig:scattering}\textcolor{blue}{(a)}) \cite{thz_rappaport}. This results in many scattered waves in addition to the primary reflected specular component, resulting in multiple signal paths across different directions. This is studied using the directive scattering (DS) model in \cite{ds_scatter_rappaport}. Results show that the scattered power is higher at higher frequencies relative to the specularly reflected power. Interestingly, the strongest scattering happens when the wave hits the surface straight on. However, when the wave grazes the surface, the scattering sharply decreases, and most of the wave's energy is reflected instead \cite{ds_scatter_rappaport}.  

THz links also suffer from high absorption upon reflection. This is demonstrated in \cite{scatter_meas} using the setup in Fig.~\ref{fig:scattering}\textcolor{blue}{(b)}. Here, the bit error rate (BER) is measured at 300~GHz for a 2-meter link involving reflection from a cinderblock wall, which exhibits both absorption and scattering. The experiment is repeated with reflection from the wall coated with metal foil (which eliminates absorption) and finally from a smooth metal plate (which eliminates both absorption and diffuse scattering). The BER is plotted against input power for three cases in Fig.~\ref{fig:scattering}\textcolor{blue}{(c)}. It can be observed that the smooth metal plate exhibits the best BER for a given power. The curve shifts right for the metal foil case, demonstrating the SNR degradation from diffuse scattering. The curve shifts significantly in the cinderblock case, suggesting high absorption.

\subsection{Antenna Considerations}
In order to compensate for the increased path loss and attenuation at THz frequencies, highly directive antennas are often required and this naturally results in a communication link which is highly directional. 
This contrasts with lower frequencies, where conventional cellular antennas can broadcast over a wide coverage angle of 120\textsuperscript{$\circ$} \cite{rappaport_5G, rodwell2013sub}.
This stark contrast in the directionality of transmission introduces noteworthy challenges at high carrier frequencies, as already observed in mm-Wave 5G.

A large-scale phased array may be used to realize high directivity. 
A linear $N$-element half-wavelength spaced antenna array experiences an increase in directivity by $10\log_{10}(N)$ \cite{balanis2016antenna}. 
This can be used to create high-directive beams. 
Using a phased array architecture also makes it possible to have electronic beam steering, which is helpful in THz communication, where the channel is highly prone to blockage. 
Using a multi-antenna architecture also makes it possible to perform MIMO techniques such as spatial multiplexing, potentially across users \cite{thz_mimo}.
It should be noted that beamforming should ideally be realized using true-time delays rather than phase shifters to avoid beam squint, which can become prominent in THz communication, where channels may consume a large fractional bandwidth \cite{true_time}. 
MEMS-based mechanical steering is also being actively investigated as another possible solution \cite{beamsteering}, though this would likely be most useful for point-to-point links where channel variations are  slow. 

High directivity can also be realized using passive structures such as a lens antenna or a parabolic reflector \cite{antenna_review}. This approach becomes advantageous at THz frequencies since higher frequencies result in greater antenna directivity, given a specific aperture area \cite{balanis2016antenna}. For an antenna with an aperture area of $A_\mathrm{e}$, the antenna gain, $G_\mathrm{a}$, is given by (\ref{antenna_gain}).
\begin{equation}\label{antenna_gain}
G_\mathrm{a}= \frac{4 \pi A_\mathrm{e}}{ \lambda^2}
\end{equation}
This equation shows that the antenna gain increases with frequency for a given aperture. Using such an antenna at the transmitter and receiver, the overall channel loss, $L$, is given by (\ref{new_loss}) after embedding the antenna gain. 
\begin{equation}\label{new_loss}
L = \left(\frac{\lambda d}{A_\mathrm{e}}\right)^2
\end{equation}
It can be observed that the channel loss now decreases with an increase in frequency. This is plotted in Fig.~\ref{fig:fspl}\textcolor{blue}{(b)} using the solid lines for an $A_\mathrm{e}$ of $2$ cm\textsuperscript{2}. THz links can thus outperform RF links regarding channel loss when using a fixed aperture antenna at the transmitter and receiver. 

When utilizing a large-scale array or passive reflector, it is essential to consider the impact on antenna far-field distance \cite{balanis2016antenna}. 
The minimum distance to the far-field region, $d_{\mathrm{far}}$, for an antenna with a maximum dimension of $D$ and wavelength of $\lambda$ is given by (\ref{far_field}). 
\begin{equation}
\label{far_field}
 d_{\mathrm{far}}= \frac{2D^2}{\lambda}
 \end{equation} 
Due to the small wavelengths, the far-field distance can become quite substantial at THz frequencies. 
For instance, employing a 10 cm antenna array can result in a far-field distance of 20 m at 300~GHz. 
This sizable distance indicates that the receiver can likely operate in the near-field, necessitating a redesign of conventional beam-steering algorithms \cite{jornet_nature_survey}. For example, \cite{airy} proposes performing wavefront engineering for link maintenance, where so-called `airy' beams are created to curve around potential blockages in the near-field.

It is worth noting that, while highly directional beams may indeed be formed at THz using the aforementioned methods, a key challenge remains in efficiently closing and reliably maintaining the link between two devices \cite{ethan_beam}. 
The overhead of traditional mm-Wave beam sweeping approaches to close the link may be unsuitable at THz, where beams are even more narrow and where path loss is even more severe. 
This necessitates new mechanisms and procedures for beam alignment, in order to ensure that communication between two devices can be established and maintained, even in mobile applications and when the environment is highly variable. 

\section{Device Technology}
    
This section compares key semiconductor technologies suitable for designing circuits in the IEEE 802.15.3d band. `Technology' or a 'process' in this context refers to a specific semiconductor manufacturing process and its typical feature size. Examples of this include silicon CMOS (complementary metal-oxide-semiconductor), SiGe BiCMOS (Silicon-Germanium Bipolar CMOS), and III-V technologies like InP (Indium Phosphide) and GaAs (Gallium Arsenide) \cite{dev_sengupta_nature}.

\subsection{Metrics for Performance Characterization}

A key challenge in THz circuit design stems from the limitations in transistor performance \cite{dev_sengupta}. \textcolor{black}{At these frequencies, transistor performance degrades to the point where it can no longer provide amplification}. Furthermore, the interconnects that allow for electrical connections to the transistor add significant parasitic resistance, capacitance, and inductance, further damaging overall performance. Transistors also exhibit non-quasistatic behavior at THz frequencies, making their modeling complicated.

\textcolor{black}{A technology-agnostic metric commonly used to characterize the transistor performance at high frequencies is the maximum oscillation frequency, $f_\mathrm{max}$. For any 2-port (such as a transistor), $f_\mathrm{max}$ is the frequency at which the maximum available power gain (or the unilateral power gain) drops to zero \cite{pozar}. In other words, amplifiers and oscillators cannot be built beyond the $f_\mathrm{max}$ of a technology.}

\textcolor{black}{Circuit design becomes challenging beyond $f_\mathrm{max}$/2. Amplifiers operating at these frequencies require several amplification stages. Consequently, power amplifiers at these frequencies have low saturated output power and low power-added efficiency, and low-noise amplifiers have high noise figures, both detrimental to RF transceiver performance. A higher $f_\mathrm{max}$ almost always implies a better communication link for the same DC power, enabling more energy-efficient communication. It should be noted that while amplification is not possible beyond  $f_\mathrm{max}$, it is possible to generate signals using non-linear circuit design techniques. However, these techniques usually have low efficiencies (\textless 0.5 \%) \cite{swami_dev}. This will be discussed later in Section VI.}

\textcolor{black}{The transit frequency, $f_\mathrm{T}$, is another metric for high-frequency technology characterization. It is the frequency at which the current gain of a transistor, with source and drain shorted, drops to unity. While the relative importance of the two parameters is debated, $f_\mathrm{max}$ is often seen as a better metric for high-frequency characterization as it accounts for more non-idealities \cite{niknejad_dev}. Hence, throughout this survey, we will emphasize $f_\mathrm{max}$ as a standard for comparing technologies.}

 \begin{figure}
        \centering
        \includegraphics[page = 5, trim=0cm 8cm 6cm 8.2cm, clip=true, width=0.45   \textwidth]{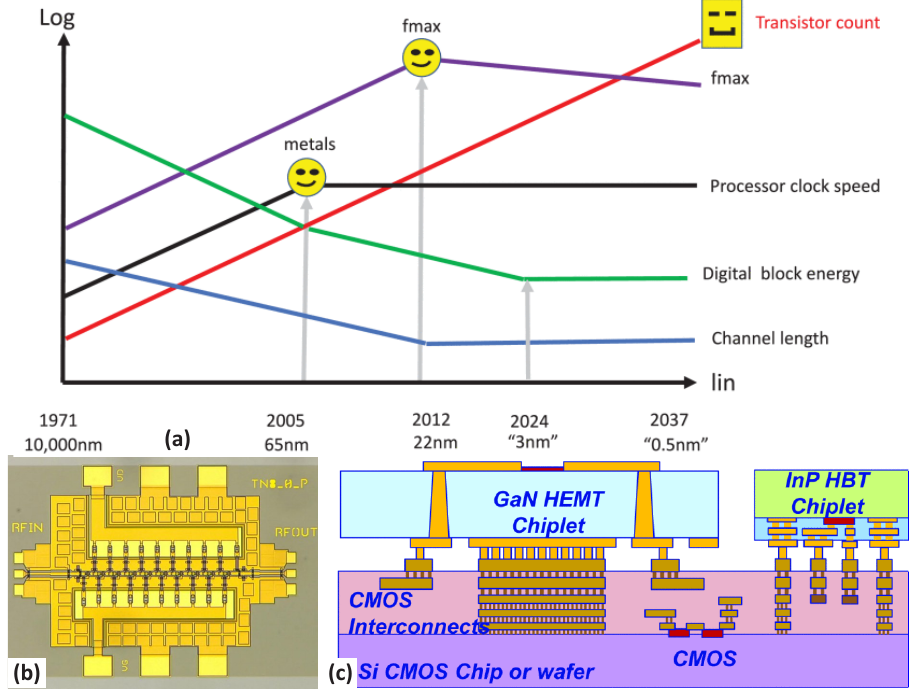}         
        \caption{ (a) Illustration of CMOS scaling, with asymptotic development following Moore's law. While the transistor count has remained increasing over the years, $f_\mathrm{max}$ peaked at 22nm \cite{nauta}. (b) First demonstration of a 1~THz amplifier using InP \cite{inp_3}. (c) Example demonstrating `best junction for the function' through heterogeneous integration \cite{hetero_1}.  }
        \label{fig:device}
 \end{figure}
 
\subsection{Silicon CMOS Technology}
Silicon CMOS is usually the most popular choice when designing integrated circuits. This is mainly due to its low cost and high digital integration capabilities, enabling systems-on-chip (SoCs) with enhanced sensing and computing capabilities. \textcolor{black}{A CMOS technology node is characterized by its feature size, which roughly describes the smallest dimension of the transistor gate length. For example, a 65nm CMOS node offers CMOS transistors with a minimum gate length of 65nm. Note that this naming convention is not strictly followed in recent technology nodes.}

CMOS technology has rapidly scaled to lower dimensions for the past several decades, following Moore's law.  This has resulted in an exponentially growing computing power. However, CMOS scaling is driven by a desire to enhance digital CMOS performance. This does not necessarily translate to enhanced RF performance \cite{nauta}. This behavior is illustrated in Fig.~\ref{fig:device}\textcolor{blue}{(a)}, where different transistor performance metrics are plotted for various technology nodes. \textcolor{black}{Notice that the transistor count has increased exponentially and is accompanied by a reduction in digital block energy, pushing Moore's law forward. However, $f_\mathrm{max}$ reached its peak at 22nm, with an $f_\mathrm{max}$ of 370 GHz \cite{GF_FDSOI}. This trend is concerning for THz circuit designers since they can no longer rely on technology scaling to push to higher frequencies. This demands new circuit design techniques or switching to a non-CMOS platform. }

\subsection{SiGe BiCMOS Technology}
The SiGe BiCMOS process is another popular choice for THz circuit design, as it provides high-performance HBTs (heterojunction bipolar transistors). These processes often utilize the same platform as a CMOS node, thereby inheriting the benefits associated with digital CMOS technology. Cutting-edge SiGe processes have achieved an $f_\mathrm{max}$ reaching up to 720 GHz \cite{hbt_ihp}. However, it is essential to acknowledge that the CMOS transistors integrated within these nodes are typically from earlier generations. This makes it difficult to integrate complex digital processing into the same chip. For instance, the most advanced BiCMOS process currently offers CMOS transistors with a 45nm feature size, a technology that is over a decade old \cite{hbt_gf}.

\subsection{III-V Technologies}

III-V semiconductors, such as gallium arsenide (GaAs) and indium phosphide (InP), present numerous benefits over conventional silicon-based technologies. These advantages stem from higher electron mobility, higher breakdown voltages, and improved high-temperature performance. InP HBTs and HEMTs (high-electron-mobility transistors) are among the favored III-V technologies for THz circuit design, capable of reaching an $f_\mathrm{max}$ beyond 1 THz \cite{inp_1,inp_2}. Northrop Grumman has showcased a 1 THz amplifier, as detailed in \cite{inp_3} (Fig.~\ref{fig:device}\textcolor{blue}{(b)}). Despite these advantages, III-V technologies have not been as widely adopted as CMOS and SiGe due to their higher costs, lack of digital integration capabilities, and smaller available wafer sizes \cite{dev_sengupta}. However, with the advent of 6G technology, III-V semiconductor \textcolor{black}{technology could significantly grow in demand and maturity}, addressing some of these challenges.

\subsection{Future Prospects}

Significant research efforts are underway to enhance device performance, highlighted by initiatives like DARPA's T-MUSIC program in North America \cite{tmusic} and the TARANTO project in Europe \cite{taranto}. For circuit design in the THz band, adopting a ``best junction for the function" strategy is often recommended (Fig.~\ref{fig:device}\textcolor{blue}{(c)}) \cite{hetero_1}. This strategy integrates advanced III-V technologies for the RF front-end with CMOS technology for digital components. Achieving this integration depends on advancements in packaging techniques and heterogeneous integration \cite{hetero_2, hetero_3}.

Resonant tunneling diodes \cite{rtd}, traveling wave tubes \cite{twt}, and photonic techniques like quantum cascade lasers \cite{qcl_ben} have also been explored for generating THz signals. These advanced technologies offer promising avenues for THz applications but face high costs, complex fabrication processes, and integration and scalability issues \cite{dev_sengupta_nature}. Despite these challenges, ongoing research aims to improve their accessibility and compatibility with existing semiconductor processes, potentially paving the way for their increased use in future THz systems.
    
\section{Circuit Design}
    Circuit design techniques at THz frequencies differ notably from those used for RF and mm-Wave frequencies. This is primarily due to transistor performance degradation from the limited $f_\mathrm{max}$ \cite{sorin_book}. \textcolor{black}{ Additionally, passive devices such as capacitors and inductors suffer from a high loss at these frequencies due to skin effect, self-resonance, and substrate coupling. Due to these limitations, designing amplifiers, local oscillators (LOs), and frequency synthesizers at THz frequencies is difficult.}

 \begin{figure}
        \centering
        \includegraphics[page = 11, trim=0cm 11cm 5.8cm 11cm, clip=true, width=0.45   \textwidth]{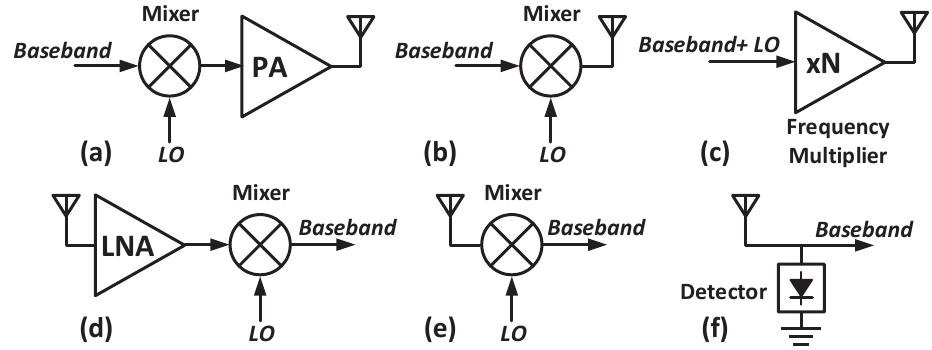}         
        \caption{Transceiver architectures suitable at THz frequencies (a) PA-last transmitter (b) Mixer-last transmitter (c) Multiplier-last transmitter (d) LNA-first receiver (e) Mixer-first receiver (f) Power detector based receiver  }
        \label{fig:circuit}
 \end{figure}

Harmonic techniques play a significant role in circuit design at these frequencies. \textcolor{black}{Any non-linear system generates higher-order harmonic products when driven by a large-signal input. Conventionally, these harmonic products are undesired and need to be removed. However, these harmonics can prove useful at THz circuit design to generate signals beyond $f_\mathrm{max}$. The greater the non-linearity, the stronger the higher-order harmonics, resulting in greater THz signal efficiency.}

Consider a non-linear device, such as a transistor, driven by an RF source. Strong harmonic products are generated if the source power is large enough to drive this device into non-linear operation. When an NMOS transistor is driven by a voltage $V_\mathrm{in}$ at frequency $f_\mathrm{0}$, the drain current $I$ can be expressed as

\begin{equation}\label{harmonics}
I = I_\mathrm{0} + g_{m1}V_\mathrm{in} + g_{m2}V_\mathrm{in}^2 + g_{m3}V_\mathrm{in}^3 + ... \;,
\end{equation}

\noindent where $ I_0$ represents the DC, while $ g_{m1}V_\mathrm{in}$ represents the output current at $ f_0$ \cite{momeni_travelling}. The output current includes higher-order terms generated due to transistor non-linearity, which are represented by $g_{m2}V_\mathrm{in}^2$ and $g_{m3}V_\mathrm{in}^3$. These terms contain harmonic signals at $2f_\mathrm0$ and $3f_\mathrm0$, respectively. These higher-order harmonic terms can be extracted and used to design THz signal sources. \textcolor{black}{Note that while non-linearity can be used to design signal sources and synthesizers beyond $f_\mathrm{max}$, it still cannot achieve amplification.}

Extensive research is underway within the circuit design community to improve the performance of transmitters, receivers, and local oscillators for THz operation. This section overviews various transmitter and receiver architectures for THz design. \textcolor{black}{Understanding trade-offs within various architectures is crucial in choosing the correct modulation schemes to maximize data rates and spectral efficiency.}

\subsection{Transmitter Architectures }

\subsubsection{Power Amplifier (PA)-Last Architecture}
This is the conventional transmitter architecture, which is followed at RF frequencies. This architecture can support both amplitude and phase modulation schemes by incorporating a linear power amplifier as its final stage, enabling higher-order modulation schemes like $M$-QAM (Fig.~\ref{fig:circuit}\textcolor{blue}{(a)}). However, due to the limited $f_\mathrm{max}$, several amplification stages are needed to achieve sufficient amplification at THz frequencies. This impacts power consumption, amplifier stability, and efficiency \cite{okada_2d}. For instance, \cite{kko_amp} demonstrates a 290 GHz power amplifier in a 130nm BiCMOS process with a saturated output power of 7.5 dBm and a power added efficiency of just 0.39\%. Because of this, PA-last techniques are usually avoided while designing in CMOS/SiGe platforms, which have a lower $f_\mathrm{max}$. However, this architecture remains popular in InP transmitters \cite{inp_transceiver}, \textcolor{black}{which have relatively high $f_\mathrm{max}$}. 

\subsubsection{Mixer-Last Architecture}
\textcolor{black}{This architecture does not use a power amplifier. Instead, the final stage here is a mixer (Fig.~\ref{fig:circuit}\textcolor{blue}{(b)})}. A mixer behaves linearly for the input baseband signal but has non-linear behavior with the LO \cite{okada_300} \textcolor{black}{and thus performs upconversion. Conventionally, the baseband signal is upconverted to the LO frequency. 
However, an $N$-th harmonic mixer can be designed, which upconverts the baseband signal to the  $N$-th harmonic of the LO.} 
This approach is capable of operating at frequencies beyond $f_\mathrm{max}$ and has been utilized in \cite{fuji_2} and \cite{fuji_1} with the second and third harmonics of the LO, respectively. 
This topology can support $M$-QAM modulation as the mixer behaves linearly and maintains the amplitude and phase information of the input baseband signal. 
However, mixers typically provide low-output saturated power.
\textcolor{black}{This affects the link budget and can thus limit the modulation order and viable link distance.} 

\subsubsection{Multiplier-Last Architecture}
This architecture utilizes a frequency multiplier as the final stage (Fig.~\ref{fig:circuit}\textcolor{blue}{(c)}). \textcolor{black}{A frequency multiplier is a circuit designed to maximize non-linearity to achieve high output power at the harmonic of the input. 
Frequency multipliers typically can achieve higher output power than a mixer and have been used in THz transceiver design} \cite{niknejad_thyagarajan_tx, reynart_outphasing, 400g_thomas}. 
However, this technique has a few noteworthy challenges. 

\begin{figure}
        \centering
        \includegraphics[page = 13, trim=0cm 9cm 10cm 9cm, clip=true, width=0.45   \textwidth]{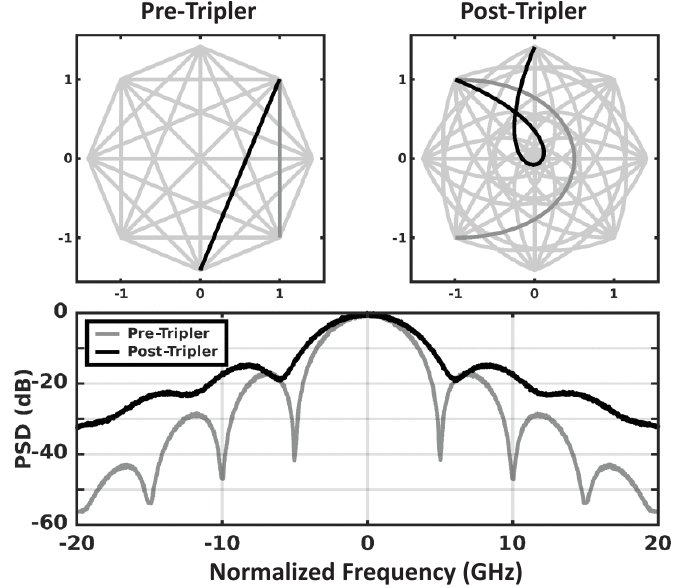}         
        \caption{Constellation diagrams and spectrum for an 8PSK signal pre-and post-tripling \cite{reynart_outphasing}. Constellation points change their position post-tripling. However, they do not fall over each other and cause signal corruption. In the frequency domain, bandwidth expansion takes place post-tripling. }
        \label{fig:multiplier}
 \end{figure}

Firstly, amplitude distortion can occur due to the inherently non-linear nature of frequency multiplication. This makes it difficult to support amplitude modulation beyond two levels. Secondly, phase distortion can also occur during frequency multiplication, causing constellation points to rotate. For instance, an 8PSK signal upon doubling would result in constellation points corrupting each other as different input symbols get mapped to the same symbol after doubling.  However, 8PSK can be preserved by tripling. This is illustrated in Fig.~\ref{fig:multiplier} \cite{reynart_outphasing}. Although the constellation points change their position post-tripling, they do not cause signal corruption. Thus, only certain modulation schemes can be supported with specific multiplication indices in a multiplier-last transmitter \cite{niknejad_thyagarajan_tx}. Lastly, bandwidth expansion is a concern due to the non-linear frequency translation in a frequency multiplier (Fig.~\ref{fig:multiplier}). This can cause the bandwidth occupied by the input baseband signal to expand by $N$ times when passed through a multiply-by-$N$ circuit \cite{reynart_outphasing}. 

It should be noted that the multiplier-last architectures can support 2-level amplitude modulation (such as OOK) without any of the problems mentioned above. This is utilized in \cite{400g_thomas} to create an efficient THz link at 400 GHz, \textcolor{black}{for instance}. 

\subsection{Receiver Architectures}

At THz frequencies, there exist three commonly used receiver architectures: the low-noise amplifier (LNA)-first architecture (Fig.~\ref{fig:circuit}\textcolor{blue}{(d)}) \cite{inp_transceiver}, the mixer-first architecture (Fig.~\ref{fig:circuit}\textcolor{blue}{(e)}) \cite{okada_300, fuji_2, fuji_3}, and the power-detector-based architecture (Fig.~\ref{fig:circuit}\textcolor{blue}{(f)} \cite{heydari_210}. Both LNA-first and mixer-first architectures share similar trade-offs with PA-last and mixer-last architectures. As they are linear, these architectures can accommodate $M$-QAM modulation schemes. Conversely, the power-detector-based architecture offers a straightforward method for amplitude demodulation while consuming low DC power. Additionally, it can be utilized for phase demodulation by applying a Kramers-Kronig processing technique, as exemplified in \cite{sengupta_kramer}. \textcolor{black}{In terms of noise figure, LNA-first architectures offer the lowest, followed by mixer-first architectures. Power-detector-based approaches typically have high noise figures (often exceeding 30 dB) as the conversion gain depends on input received power. }

\subsection{Summarizing Remarks}
The choice of the best THz transceiver architecture is still debated. However, some conclusions can be drawn based on the discussions above. 
\begin{itemize}
    \item If a technology node with a high $f_\mathrm{max}$ is used for design (such as III-V), then a PA-last and LNA-first transceiver, with $M$-QAM modulation is optimal.
    \item  If one desires to perform $M$-QAM modulation in a technology with limited $f_\mathrm{max}$ (such as CMOS, SiGe), then mixer-last and mixer-first approaches should be used. While this can support high data rates, the communication link distance can be limited since this approach suffers from low saturated output power and a high noise figure. 
    \item  If the demand is for a robust link with moderate data rates in a technology with limited $f_\mathrm{max}$ (such as CMOS, SiGe), then the multiplier last approach is optimal, as it can support high output power. On the receiver end, a mixer-first architecture may be optimal due to its lower noise figure compared to the detector-first approach.
\end{itemize}

 \section{Antennas and Packaging 
 }
    A 300~GHz signal has a wavelength of 1 mm in free space. This enables the design of large antenna arrays within a small area. Moreover,  this wavelength decreases in a dielectric medium. For example, in silicon dioxide—a widely used dielectric in technologies like silicon CMOS and SiGe BiCMOS—the wavelength of a 300~GHz signal is reduced to just 0.5 mm due to a dielectric constant of 3.9 \cite{lens_pack_4}. This reduction enables the design of a large antenna array within an integrated circuit (IC). For instance, a 144-element on-chip array is demonstrated at 670 GHz in \cite{ant1}.

Efficient antennas and their interfacing methods are crucial. In an RF signal chain, the LNA and PA interface with the antenna. Inefficiencies in the antenna or its interfacing can lead to increased noise figure in the LNA and decreased efficiency and output power in the PA, highlighting the importance of developing high-efficiency antennas and employing low-loss packaging techniques for interfacing \cite{wv_package_1}. Besides \textcolor{black}{this}, packages also need to provide mechanical support and protection from external conditions and facilitate easy thermal dissipation while enabling easy interfaces with DC biases and I/O control signals.  

This section explores the latest advancements in antenna design and packaging approaches at THz frequencies. While our primary emphasis is silicon ICs, the techniques and principles discussed apply to a broad spectrum of other technologies.

 \begin{figure*}
        \centering
        \includegraphics[page = 6, trim=0cm 8cm 0cm 8cm, clip=true, width=0.95   \textwidth]{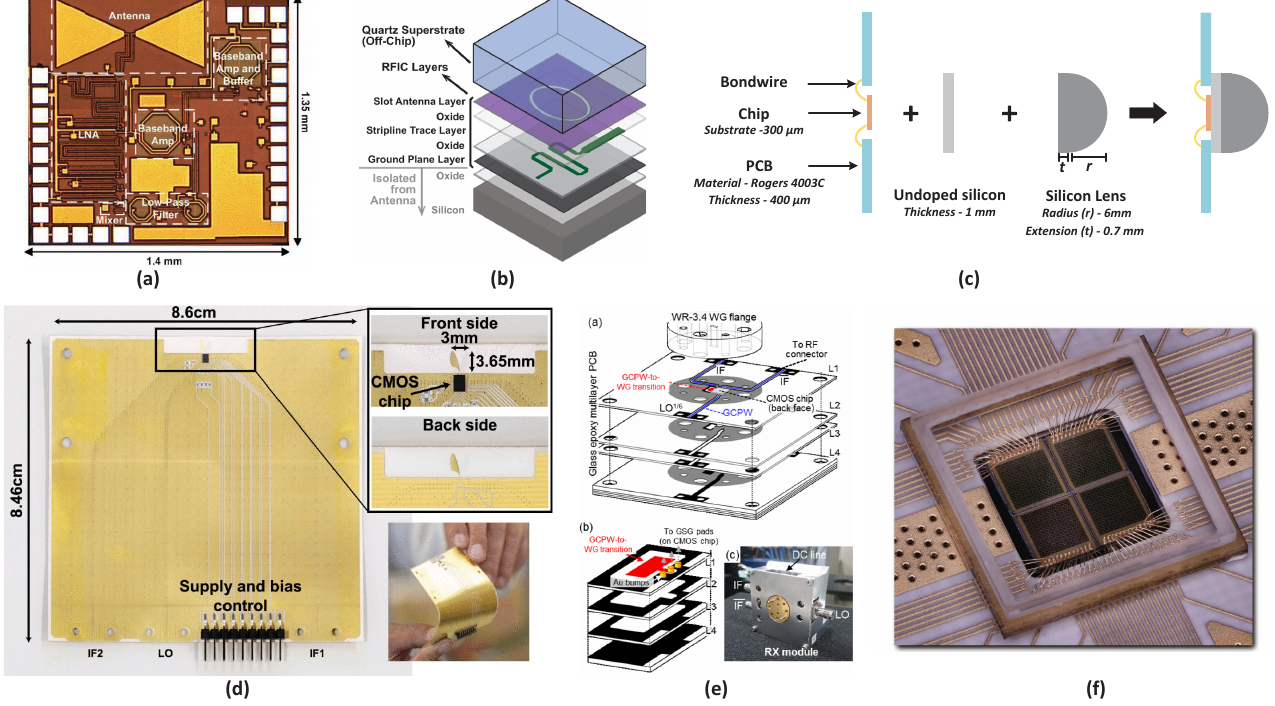}             
        \caption{(a) An on-chip bowtie antenna implemented in a 130nm SiGe BiCMOS platform \cite{lens_pack_3}. (b) Adding a quartz superstrate to increase the directivity and gain of top side radiation \cite{lens_pack_2}. (c) Silicon lens packaged to a silicon die \cite{400g_thomas}. (d)  Photographs of the PCB with Vivaldi antennas used for a 300~GHz phased-array \cite{okada_300}. (e) CMOS chip to a waveguide transition built on a multilayered glass epoxy PCB \cite{wv_package_5}. (f) Four CMOS chips (each measuring 2 mm × 2 m) consisting of a programmable two-dimensional array of 12 × 12 meta-elements \cite{ris_1}. }
        \label{fig:antenna}
 \end{figure*}

\subsection{Integrated On-Chip Antennas}
Given the reduced wavelength in silicon, it is feasible to integrate antennas directly within integrated circuits (\textcolor{black}{ICs} typically have a size of a few millimeters). These antennas are often designed in thick top metal layers to mitigate conductive losses. A variety of antenna designs—including dipole, slot, ring, cavity, leaky-wave, and patch antennas—have been successfully implemented on-chip \cite{wv_package_3, wv_package_4, lens_pack_2, ant2}. For example, Fig.~\ref{fig:antenna}\textcolor{blue}{(a)} demonstrates an on-chip bowtie antenna implemented in a 130nm SiGe BiCMOS platform. \cite{lens_pack_3}. These antennas can be designed to radiate from the top or back sides of the IC, leading to different design strategies, \textcolor{black}{which are discussed below}.

\subsubsection{Top-Side Radiation}
Antennas that radiate from the top typically exhibit lower efficiencies. This is because of the significant mismatch in intrinsic impedances between the antenna medium (silicon dioxide) and air \cite{lens_pack_4}. An effective strategy to enhance radiation efficiency involves the addition of a quartz superstrate atop the antenna, a technique that has demonstrated improved radiation efficiency \cite{lens_pack_2}. This is demonstrated in Fig.~\ref{fig:antenna}\textcolor{blue}{(b)}. While this method does necessitate some post-fabrication processing, the associated costs are relatively low.

\subsubsection{Back-Side Radiation}
Radiation from the back side generally achieves higher efficiency compared to top-side radiation. This improvement is attributable to the high dielectric constant of the silicon substrate beneath the silicon dioxide layer, which lowers intrinsic impedance. However, back-radiation faces challenges, as unwanted surface waves can get generated within the silicon substrate. To address this, a hyper-hemispherical silicon lens can be added, which can significantly boost efficiency and directivity \cite{lens_pack_4, lens_pack_1, 400g_thomas}. This is demonstrated in Fig.~\ref{fig:antenna}\textcolor{blue}{(c)}. Yet, adopting silicon lenses has drawbacks: they are costly and require precise alignment. When utilized with an array, lenses typically suffer from poor performance due to off-axis effects, where the phase center of individual antenna elements does not align with the lens's phase center \cite{ant3}.

Recent advancements have seen the use of 3D-printed Teflon lenses, which offer enhanced directivity owing to the design flexibility afforded by 3D printing in creating optimal lens shapes \cite{ant4}. Additionally, the exploration of metasurface-based planar lenses represents a growing field of study \cite{ant5}.

\subsubsection{Antenna on PCB}
Antennas can be integrated onto printed circuit boards (PCBs), offering a cost-effective and flexible solution \cite{pcb_1, okada_300}. Fig.~\ref{fig:antenna}\textcolor{blue}{(d)} demonstrates a Vivaldi antenna array at 300~GHz, implemented by stacking PCBs over one another. However, this approach has notable drawbacks. First, connecting the chip to the PCB requires either bond wires or flip-chip packaging using copper bumps. Both methods can adversely affect impedance matching and increase signal loss. Second, the material typically used in PCBs exhibits high loss at THz frequencies, potentially lowering antenna efficiency. Additionally, the manufacturing resolution available at most PCB facilities may not meet the precise requirements for designing antennas at these high frequencies, particularly for those with sub-millimeter dimensions.

\subsubsection{Micro-Machined Waveguide Antennas}
Waveguide antennas are known for their exceptional performance at THz frequencies. However, interfacing these antennas with ICs presents technical challenges. 
Typically, bond wires or copper bumps are used to deliver the THz signal to an EM coupler, which then excites the waveguide antenna.
This is demonstrated in Fig.~\ref{fig:antenna}\textcolor{blue}{(e)}. 
However, achieving consistent impedance matching proves complex in this approach \cite{wv_package_2, wv_package_5}. 
An alternative strategy involves embedding the EM coupler directly within the IC, enabling it to directly generate the required EM mode for the antenna \cite{wv_package_3, wv_package_4}. 
Despite the high quality of antennas and interfaces this method yields, it is expensive due to the high costs involved in micro-machining at such fine dimensions.

\subsubsection{Other Techniques}
Dielectric waveguides (DWG) have attracted attention recently, where a low-loss polymeric medium is used to guide electromagnetic waves at \textcolor{black}{THz frequencies. 
DWGs exhibit low loss and can hence potentially replace fiber in medium-range links \cite{dwg_01, dwg_02}}. 
However, it is important to note that DWGs are not strictly a wireless technology \textcolor{black}{since it requires a dielectric channel to guide the EM wave}. 

Another area of significant interest is the development of reconfigurable intelligent surfaces (RIS). Designing these advanced structures to operate at THz frequencies presents substantial challenges, primarily because conventional switches fail to function at these high frequencies, complicating reconfigurability. Despite this, there have been demonstrations of THz RIS structures such as metasurfaces for holographic projections \cite{ris_1} (Fig.~\ref{fig:antenna}\textcolor{blue}{(f)}) and reflectarrays using 1-bit phase shifter for high-resolution radar \cite{ris_2}.

\section{\textcolor{black}{THz Transceiver Demonstrations}}
    The focus on THz frequencies has been driven by their largely untapped potential, evidenced by practical applications such as the use of frequencies around 120~GHz for data transmission during the Beijing Olympics \cite{beijing} and the Air Force Research Laboratory's (AFRL) experiments on propagation loss between aircrafts at frequencies of 300~GHz \cite{afrl}. These applications, among others, underscore the significant interest and ongoing efforts within the field to overcome the inherent challenges of operating at such high frequencies. This section delves into a few \textcolor{black}{prototype transceivers above 200~GHz} developed by academic and industrial research institutions. We highlight some notable demos that have successfully demonstrated a complete over-the-air (OTA) wireless link, encompassing both transmitter and receiver components, capable of achieving multi-Gbps transmission rates. 

 \begin{figure*}
        \centering
        \includegraphics[page = 10, trim=0cm 8.5cm 0cm 8.6cm, clip=true, width=0.95   \textwidth]{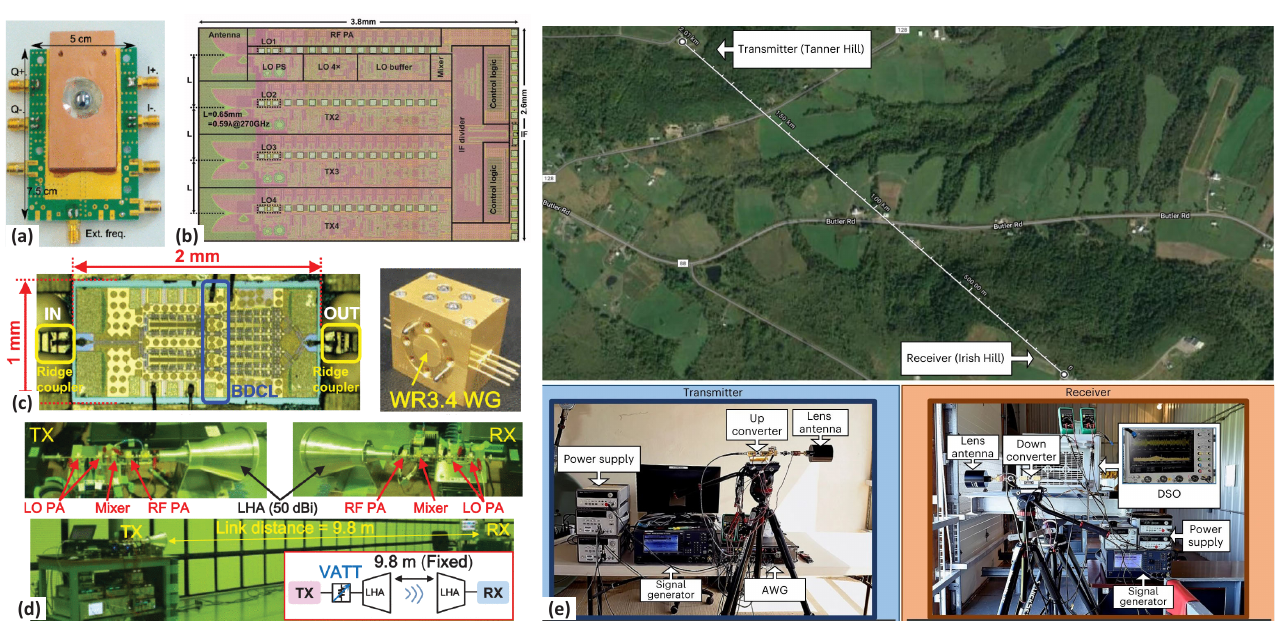}         
        \caption{(a) 230 GHz BiCMOS transceiver from the University of Wuppertal. The silicon lens and heat-sinking can be observed here \cite{pfieffer}. (b) 300~GHz phased array transmitter, with on-chip Vivaldi antenna array IC from TokyoTech \cite{okada_2d}. (c) InP power amplifier chip from NTT and TokyoTech with a ridge-coupler and waveguide packaging \cite{inp_transceiver} (d) 300~GHz InP transceiver achieving a 9.8m OTA link \cite{inp_transceiver} (e) A 2 km point-to-point link at the Air Force Research Laboratory research facility, and the corresponding transmitter and receiver hardware \cite{jornet_km}.
}
        \label{fig:demos}
 \end{figure*}
 
\subsection{Hiroshima University and NTT}

Hiroshima University and Nippon Telegraph and Telephone (NTT) have showcased a series of transceivers within the 300~GHz band, using a 40nm CMOS technology platform \cite{fuji_1, fuji_2, fuji_3}. Notably, \cite{fuji_2} presents a complete transceiver capable of sustaining a 3~cm over-the-air (OTA) link with a data rate of 80~Gbps while achieving an energy efficiency of 22.3~pJ/bit. This transceiver supports advanced modulation schemes, including QPSK, 16QAM, and 64QAM. Note that this prototype does not include a packaged antenna and interfacing. Instead, the OTA link was tested by interfacing the chip with external waveguide horn antennas via waveguide probes.

Further advancements \textcolor{black}{by the same group} are detailed in \cite{fuji_3}, where the authors demonstrate a receiver module that incorporates packaging. In this development, the 40nm CMOS chip is flip-chip bonded onto a PCB and interfaces with waveguide antennas (Fig.~\ref{fig:antenna}\textcolor{blue}{(e)}). This packaged receiver module can achieve 76~Gbps using 16QAM modulation over a 6~cm distance and 4.32~Gbps with QPSK modulation over a 1~meter OTA distance.

\subsection{BiCMOS Transceiver from University of Wuppertal}
The study referenced as \cite{pfieffer} showcases a BiCMOS transceiver operating at 230~GHz. The transceiver is equipped with on-chip ring antennas and a silicon lens for optimal radiation, while heat-sinking techniques are implemented to regulate thermal performance (Fig.~\ref{fig:demos}\textcolor{blue}{(a)}). Notably, this transceiver can attain a data transmission rate of 100~Gbps using 16QAM modulation over a 1-meter link while maintaining an energy efficiency of 14~pJ/bit.

\subsection{Beam-Steerable Phased Array from TokyoTech}
TokyoTech has successfully showcased beam steering within the 300~GHz band (from 242-280~GHz), utilizing a 65nm CMOS technology platform \cite{okada_300}. The prototype features a CMOS transceiver paired with PCB-based Vivaldi antennas and can support QPSK and 16QAM modulation schemes (Fig.~\ref{fig:antenna}\textcolor{blue}{(d)}). By stacking four of these PCBs, the team has managed to steer the beam across a 36-degree angle. Furthermore, the design can also operate with standard horn antennas. OTA tests have resulted in data transmission speeds of 16 Gbps over a distance of 3.5~cm while achieving an energy efficiency of 93.75~pJ/bit. Additionally, the team recently demonstrated a new 2-D phased array transmitter by stacking multiple PCBs and chips with on-chip Vivaldi antennas \cite{okada_2d} (Fig.~\ref{fig:demos}\textcolor{blue}{(b)}). 

\subsection{NTT and TokyoTech}
NTT and TokyoTech have demonstrated a 300~GHz heterodyne transceiver using their in-house 80nm InP HEMT platform \cite{inp_transceiver}. It consists of custom-designed PAs and mixer modules, which are all packaged in individual waveguide modules by using ridge couplers (Fig.~\ref{fig:demos}\textcolor{blue}{(c)}). Using external high-gain lens antennas, the transceiver achieves OTA data transmission of a 120~Gb/s 16QAM signal over a link distance of 9.8~m (Fig.~\ref{fig:demos}\textcolor{blue}{(d)}), with an energy efficiency of 92.5~pJ/bit.

\subsection{Multi-Kilometer Link from Northeastern University}
Northeastern University has showcased a long-distance, high-speed link at 225~GHz for wireless backhaul \cite{jornet_km}. The link supports over 2~Gbps at frequencies between 210–230~GHz across a 2~km OTA outdoor environment, utilizing a 200~mW terahertz signal source and broadband low-noise balanced mixers employing Schottky diode technology developed by NASA (Fig.~\ref{fig:demos}\textcolor{blue}{(e)}). The system employs a highly directional lens antenna (with over 40~dBi gain) for the outdoor link setup. It also integrates custom communication and signal processing strategies into a software-defined, ultra-wideband (30~GHz) digital baseband system. It presents cutting-edge outcomes, proving the feasibility of establishing long-distance terahertz links in real-world outdoor settings.

 \begin{figure}
        \centering
        \includegraphics[page = 14, trim=0.2cm 12.2cm 10.1cm 12cm, clip=true, width=0.49   \textwidth]{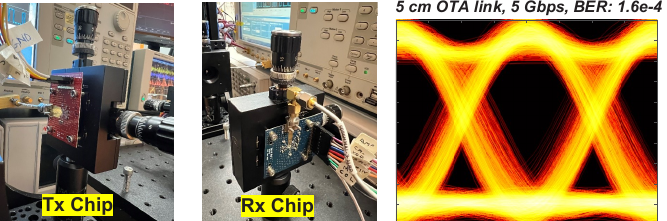}         
        \caption{\textcolor{black}{400 GHz transceiver from UCLA and measured eye-diagram for 5~Gbps transmission across a 5~cm OTA link \cite{400g_thomas}.}}
        \label{fig:ucla_400}
 \end{figure}

\subsection{400 GHz Transceiver from UCLA}
UCLA has demonstrated a multi-Gbps wireless transceiver at 400~GHz using a 90nm SiGe BiCMOS process \cite{400g_thomas}. 
This work utilizes silicon PIN diodes that show strong non-linearity and can hence generate THz signals with high DC-to-THz generation efficiency.
The prototype supports OOK modulation and incorporates on-chip antennas paired with a silicon lens. 
In OTA experiments, this prototyped transceiver achieved 5~Gb/s of data transmission over a link distance of 5~cm (Fig.~\ref{fig:ucla_400}\textcolor{blue}), with an energy efficiency of 52.8~pJ/bit. 
Notably, this is the first instance of a fully integrated multi-Gb/s wireless transceiver operating above 300~GHz in silicon. 
Additionally, by employing external mirrors for collimation, UCLA demonstrated a transmitter that can support 3~Gbps over a  20-meter link \cite{thomas_long_dist}.

\section{Conclusion}
    The potential to transform wireless communication and sensing by operating at THz carrier frequencies has spawned research and development across the globe, with eyes set on enabling future 6G networks. 
In this survey, we provide a comprehensive summary of recent advancements in THz technology, spanning state-of-the-art systems, circuits, devices, and antennas, while also calling attention to directions needing further development. 
We also highlight notable experimental demonstrations of THz technology from across the globe, which reveals outstanding progress but also noteworthy shortcomings related to energy efficiency and reliability.

Worthwhile directions for future research include developing process technologies that can provide devices with high $f_\mathrm{max}$ for the THz front-end, alongside advanced sub-micron CMOS for the digital back-end.
There is also a need for novel circuit topologies with improved efficiency and antennas that offer good performance without requiring expensive post-processing and precise machining.
Furthermore, novel algorithms are needed to facilitate reliable operation in THz channels, which are often directive, lossy, and prone to blockage.

Despite its practical hurdles, global enthusiasm for THz communication and sensing raises hope that it will indeed be a core component in revolutionizing future generations of connectivity. As THz technology continues to mature, it is crucial that we conduct thorough investigations to understand the potential long-term biological effects of THz radiation.
Additionally, it is imperative that THz systems be deployed so that they do not interfere with radio astronomy and Earth observation satellites, which monitor climate change, natural disasters, and other atmospheric phenomena. 
Furthermore, the potential for increased surveillance capabilities through THz sensing demands heightened security and privacy measures.
Finally, while THz technologies hold the potential to revolutionize internet access and connectivity, it is crucial to not worsen the existing digital divide and ensure equitable access to these breakthroughs.

\hypersetup{colorlinks=true, linkcolor=blue, citecolor=blue, urlcolor=black}




\end{document}